\documentclass[12pt]{article}
\usepackage[dvips]{graphicx}

 1
\makeatletter
\@addtoreset{equation}{section}

\makeatother
\begin{document}
\begin{flushright}
KEK-TH-1086\\
SISSA 33/2006/EP
\end{flushright}
\vspace{2mm}
%
\begin{center}
{\Large\bf 
 Low scale gravity mediation with warped extra dimension \\
\vspace*{3mm}
and collider phenomenology on the hidden sector}
\end{center}
\vspace{8mm}
\begin{center}
\normalsize
{\large \bf 
Hideo Itoh$^{(a)}$
 \footnote{E-mail: hideo@post.kek.jp},  
Nobuchika Okada$^{(a,b)}$
 \footnote{E-mail: okadan@post.kek.jp}
 and 
Toshifumi Yamashita$^{(c)}$
   \footnote{E-mail: yamasita@sissa.it}}
\end{center}
\vskip 1.2em
\begin{center}
${}^{(a)}$ {\it 
Theory Division, KEK, 
Oho 1-1, Tsukuba 305-0801, Japan } 

${}^{(b)}$ {\it 
Department of Particle and Nuclear Physics, 
The Graduate University \\ 
for Advanced Studies (Sokendai), 
Oho 1-1, Tsukuba 305-0801, Japan} 

${}^{(c)}$ {\it 
SISSA, Via Beirut 4, I-34014 Trieste, Italy 
}
\end{center}
\vskip 1.0cm
\begin{center}
{\large Abstract}
\vskip 0.7cm
\begin{minipage}[t]{16cm}
\baselineskip=19pt
\hskip4mm
We propose a scenario of gravity mediated supersymmetry breaking 
 (gravity mediation) in a supersymmetric Randall-Sundrum model. 
In our setup, both of the visible sector and 
 the hidden sector co-exist on the infrared (IR) brane. 
We introduce the Polonyi model as a simple hidden sector. 
Due to the warped metric, 
 the effective cutoff scale on the IR brane is ``warped down'', 
 so that the gravity mediation occurs at a low scale. 
As a result, the gravitino is naturally the lightest superpartner (LSP) 
 and contact interactions between the hidden and the visible 
 sector fields become stronger. 
We address phenomenologies for various IR cutoff scales. 
In particular, we investigate collider phenomenology 
 involving a scalar field (Polonyi field) in the hidden sector 
 for the case with the IR cutoff around 10 TeV. 
We find a possibility that the hidden sector scalar 
 can be produced at the LHC and 
 the International Linear Collider (ILC). 
Interestingly, the scalar behaves like 
 the Higgs boson of the standard model 
 in the production process, 
 while its decay process is quite different 
 and, once produced, it will provide us 
 with a very clean signature. 
The hidden sector may be no longer hidden. 
\end{minipage}
\end{center}
\newpage
%
\def\barr{\begin{eqnarray}}
\def\earr{\end{eqnarray}}


\section{Introduction}
Supersymmetric (SUSY) extension of the standard model is 
 one of the most promising ways 
 to solve the gauge hierarchy problem of the standard model. 
The minimal supersymmetric standard model (MSSM) 
 is the simplest supersymmetric extension of the standard model, 
 and its various phenomenological aspects 
 have been investigated for many years. 
However, since no superpartner has been observed 
 in the current experiments, 
 SUSY should be broken at low energies. 
The origin of SUSY breaking and its mediation mechanism 
 to the visible (MSSM) sector 
 is one of the most important issues 
 in any supersymmetric phenomenological models. 

To be consistent with our observations that 
 the Nature is almost flavor blind and CP invariant, 
 the way to transmit the SUSY breaking 
 to the visible sector is severely constrained. 
For a few decades, various mechanisms 
 for the SUSY breaking mediation 
 have been proposed in the context of four dimensional models 
 and also brane world scenarios \cite{Luty}. 
Each proposed model provides typical soft SUSY breaking mass spectra. 
Once superpartners are observed at future colliders 
 and their mass spectra are precisely measured, 
 the origin of the SUSY breaking mediation mechanism 
 could be revealed. 

The simplest model of SUSY breaking is the Polonyi model \cite{Polonyi}, 
 where a chiral superfield singlet under the standard model 
 gauge group and its tadpole term in superpotential 
 are introduced. 
Then, non-zero $F$-term is developed, and SUSY is broken. 
After SUSY is broken, 
 the SUSY breaking is transmitted to the visible sector 
 through some interactions such as gravity interactions 
 or gauge interactions. 
Operators relevant to the SUSY breaking mediation 
 are effectively described as higher dimensional contact operators 
 between the hidden sector and the visible sector superfields. 
The scale of the SUSY breaking mediation is characterized 
 by the mass scale of the contact operators. 
There are two well-known examples of SUSY breaking mediation. 
One is the gravity mediation 
 in the minimal supergravity scenario \cite{mSUGRA}, 
 where the scale of the SUSY breaking mediation 
 is the Planck scale 
 which is nothing but the cutoff scale of supergravity. 
The other is the gauge mediated SUSY breaking (GMSB) 
 \cite{GMSB}, where SUSY breaking is transmitted 
 through the standard model gauge interactions 
 with the so-called ``messenger'' fields. 
The scale of the gauge mediation is characterized 
 by the mass scale of the messenger fields, 
 which is far below the Planck scale.

In this paper, we propose a new scenario of the gravity mediation 
 in a supersymmetric Randall-Sundrum model. 
We introduce both of the visible and the hidden sectors 
 on the infrared (IR) brane. 
As a simple hidden sector we take the Polonyi model, 
 and consider the gravity mediation 
 through contact operators 
 between the hidden sector and the visible sector superfields. 
As first proposed by Randall and Sundrum \cite{RS}, 
 in four dimensional effective theory, 
 an original mass scale on the IR brane 
 is ``warped down'' to a low scale by the warp factor. 
Therefore, in our model, 
 the gravity mediation occurs at the low scale 
 due to the warping down of the original cutoff scale 
 of the model%
\footnote{
Here, the warp factor is not necessarily so strong 
 to solve the hierarchy problem completely. 
 The remaining hierarchy is solved by SUSY. 
}.
We call this scenario ``low scale gravity mediation''.  

As a result of the SUSY breaking mediation at the low scale, 
 the gravitino is naturally the lightest superpartner (LSP), 
 so that it can be a candidate of the dark mater 
 in the present universe. 
Recently, this LSP gravitino scenario has been intensively 
 studied in cosmology \cite{sWIMPcosmology} and 
 also in collider physics \cite{sWIMPcollider}. 
Our model can naturally provide this scenario.

Our model has further interesting features. 
The contact operators relevant to the gravity mediation 
 also provide contact interactions 
 between a scalar field (Polonyi field) in the hidden sector 
 and the standard model fields. 
In the context of the warped extra dimension, 
 the effective cutoff scale can be as low as 1 TeV 
 without any serious fine-tuning for parameters in the model. 
We will find a possibility that  
 the hidden sector scalar can be produced at the LHC and the ILC 
 with a very clean signature, 
 if the effective cutoff scale is low enough.

This paper is organized as follows. 
In the next section, we propose a SUSY model 
 with a warped extra dimension 
 which realizes the low scale gravity mediation. 
We also present a concrete model of the hidden sector as an example, 
 which is nothing but the Polonyi model on the IR brane. 
In Sec.~3, we address various phenomenological aspects of our model. 
In particular, we focus on collider phenomenologies 
 involving the hidden sector scalar, 
 and find a possibility that the hidden sector scalar 
 can be discovered at the LHC and the ILC. 
The last section is devoted to summary and discussions.

\section{Low scale gravity mediation}
We consider a SUSY model 
 in the warped five dimensional brane world scenario \cite{RS}. 
The fifth dimension is compactified on the orbifold $S^1/Z_2$ 
 with two branes, ultraviolet (UV) and infrared (IR) branes, 
 sitting on each orbifold fixed point. 
With an appropriate tuning for cosmological constants 
 in the bulk and on the branes,
 we obtain the warped metric \cite{RS}, 
\begin{eqnarray}
 d s^2 = e^{-2 k r |y|} \eta_{\mu \nu} d x^{\mu} d x^{\nu} 
 - r^2 d y^2 , 
\end{eqnarray}
 for $-\pi\leq y\leq\pi$, where $k$ is the AdS curvature, and 
 $r$ and $y$ are the radius and the angle of $S^1$, respectively. 

By the compactification on the orbifold, 
 N=1 SUSY of the five dimensional theory, 
 which corresponds to N=2 SUSY in the four dimensional point of view, 
 is broken down to four dimensional N=1 SUSY. 
Supergravity Lagrangian of this system can be described 
 in terms of the superfield formalism 
 of four dimensional N=1 SUSY theories \cite{SUSYL1, SUSYL2, SUSYL3}. 
For simplicity, here we consider only the gravity multiplet 
 in the bulk whose Lagrangian is given by 
\begin{eqnarray}
 {\cal L}_{\rm bulk} &=& 
 -  3 \int d^4 \theta  \frac{M_5^3}{k}  
 \left(\phi^\dagger \phi - \omega^\dagger \omega \right), 
 \label{pureSUGRA} 
\end{eqnarray} 
where $M_5$ is the five dimensional Planck mass, 
 $\phi=1+\theta^2 F_\phi $ is the compensating multiplet
 in the superconformal framework of supergravity 
 \cite{superconformal}, 
 and $\omega = \phi e^{-\pi k T}$ 
 with a radion chiral multiplet $T$ whose real part of 
 the scalar component is the fifth dimensional radius. 
Lagrangian for some chiral and gauge multiplets 
 on the UV brane are generally described as 
\begin{eqnarray}
 {\cal L}_{\rm UV}^{\rm chiral} 
 &=& 
 \int d^4 \theta \; \phi^\dagger \phi \; {\cal K}_{\rm UV} 
 + \left( 
 \int d^2 \theta \; \phi^3 W_{\rm UV} + {\rm h.c.} \right), 
 \nonumber \\
{\cal L}_{\rm UV}^{\rm gauge} 
 &=& 
  \frac{1}{4} 
 \int d^2 \theta \; f_a {\cal W}^{a \alpha} {\cal W}^a_{\alpha} 
 + {\rm h.c.} , 
 \label{UVLagrangian}
\end{eqnarray}  
where ${\cal K}_{\rm UV}$ and $W_{\rm UV}$ are 
 Kahler potential and superpotential, respectively, 
 and $f_a$ is the gauge kinetic function. 
Replacing $\phi$ by $\omega$ due to the warped metric, 
 we obtain general Lagrangian 
 for some chiral and gauge multiplets on the IR brane, 
\begin{eqnarray}
 {\cal L}_{\rm IR}^{\rm chiral} 
 &=& 
 \int d^4 \theta \omega^\dagger \omega {\cal K}_{\rm IR} 
 + \left( 
 \int d^2 \theta \omega^3 W_{\rm IR} + {\rm h.c.} \right), 
 \nonumber \\
{\cal L}_{\rm IR}^{\rm gauge} 
 &=& 
  \frac{1}{4} 
 \int d^2 \theta f_a {\cal W}^{a \alpha} {\cal W}^a_{\alpha} 
 + {\rm h.c.} . 
\end{eqnarray}  

The setup of our model is that 
 both of the hidden and visible sectors reside on the IR brane. 
Except for gravity multiplet residing in the bulk, 
 this is the same setup as in usual four dimensional models. 
We introduce a simple hidden sector 
 with a chiral superfield ($X$) 
 singlet under the standard model gauge group, 
 by whose $F$ component SUSY is broken. 
We set free parts 
 in the Kahler potential and the gauge kinetic functions 
 for each superfield of the canonical form 
 such as ${\cal K}_{\rm IR}^{\rm free} 
 =\sum_i Q_i^\dagger Q_i+X^\dagger X$ and $f_a^{\rm free}=1$, 
 where $Q_i$ denotes matter and Higgs multiplets in the MSSM 
 with flavor index $i$. 

Now let us consider the gravity mediation on the IR brane, 
 namely SUSY breaking is transmitted through contact operators 
 between the visible and the hidden sector superfields. 
For the gravity mediation in four dimensional models, 
 the contact operators are suppressed 
 by the four dimensional Planck mass, 
 which is nothing but the cutoff 
 of four dimensional supergravity. 
In our case, the original cutoff should be 
 the five dimensional Planck mass. 
In addition to the free parts of 
 the Kahler potential and the gauge kinetic functions, 
 we introduce the following contact operators 
 relevant to the gravity mediation, 
\begin{eqnarray} 
{\cal L}_{\rm contact} &=&  
 - \int d^4 \theta \omega^\dagger \omega \; 
  \left( 
  c_A^{ij} \frac{X+X^\dagger}{M_5} 
+ c_0^{ij} \frac{ X^\dagger X }{M_5^2} 
\right) Q_i^\dagger Q_j  \nonumber  \\
&-&  \frac{1}{4} \int d^2 \theta  \; 
 c_{a} 
 \frac{X}{M_5} {\cal W}^{a \alpha} {\cal W}^a_\alpha +{\rm h.c.} ,   
 \label{contact} 
\end{eqnarray} 
where $c^{ij}_A$, $c_0^{ij}$ and $c_a$ are dimensionless parameters 
 naturally of order one, 
 and $a=1,2,3$ corresponds to $U(1)_Y$, $SU(2)_L$ and $SU(3)_c$ 
 gauge groups of the standard model. 
Although the coefficients, $c^{ij}_A$ and $c_0^{ij}$, 
 are generally flavor dependent, 
 we assume the universal coefficients (minimal ansatz), 
 $c^{ij}_A=c_A$ and $c_0^{ij}=c_0$, 
 as usual in the minimal supergravity scenario, 
 otherwise flavor-changing-neutral-current (FCNC) processes 
 through superpartners exceed the current experimental bounds.

Note that, in the present form, superfields have not yet 
 been suitably normalized, because of the warped metric. 
The correct description in effective four dimensional theory 
 is given by replacing each chiral superfield as 
 $ Q_i, X \rightarrow Q_i/\omega, X/\omega$ 
 so as to eliminate $\omega$ from their free kinetic terms. 
Now we arrive at the contact operators 
 in effective four dimensional theory of the form, 
\begin{eqnarray} 
{\cal L}_{\rm contact}^{\rm eff} &=&  
 - \int d^4 \theta \; 
  \left( 
  c_A \frac{X+X^\dagger}{\Lambda_{\rm IR}} 
+ c_0 \frac{ X^\dagger X }{\Lambda_{\rm IR}^2} 
\right) Q_i^\dagger Q_i  \nonumber \\
&-&  \frac{1}{4} \int d^2 \theta  \; 
 c_{a} 
 \frac{X}{\Lambda_{\rm IR}} 
 {\cal W}^{a \alpha} {\cal W}^a_\alpha +{\rm h.c.}.  
 \label{contactOP} 
\end{eqnarray} 
Here, 
 the effective cutoff $\Lambda_{\rm IR} = \omega M_5$ appears. 
This is the most important feature of a model 
 with the warped extra dimension, 
 that is, any dimensional parameters on the IR brane 
 are inevitably warped down according to their mass dimensions 
 in effective four dimensional theory. 
As discussed in the original paper \cite{RS}, 
 $\Lambda_{\rm IR} \ll M_P$ can be achieved 
 with a mild hierarchy among the original parameters. 
For example, $\Lambda_{\rm IR} \sim 1$ TeV 
 can be realized by $M_5 \sim k \sim 11.3/r$. 
Here, $M_P=2.4 \times 10^{18}$ GeV is the reduced Planck mass 
 in four dimensions which is defined as 
 $M_P^2 = M_5^3/k$ in the strongly warped case $\omega \ll 1$. 
Thus, we can take any value of the IR cutoff 
 without theoretical difficulty.

Once non-zero $F$-term of the hidden sector field, $F_X$, 
 is developed, 
 the contact operators introduced above 
 lead to soft SUSY breaking terms in the visible sector. 
Assuming $ \langle X \rangle \ll \Lambda_{\rm IR}$, 
 for simplicity, 
 scalar squared masses, $A$-parameter and gaugino masses are
 extracted as 
\footnote{
In general, we can introduce higher dimensional terms 
 among $X$ and Yukawa couplings in superpotential, 
 which induce additional $A$-parameters. 
Throughout the paper, we do not consider 
 higher dimensional operators in superpotential, for simplicity. 
} 
\begin{eqnarray}
\tilde{m}^2 &=& \left( c_A^2 + c_0 \right)  
  \frac{|F_X|^2}{\Lambda_{\rm IR}^2} ,  \\
 A &=& 3\;  c_A\;  \frac{F_X}{\Lambda_{\rm IR}} , \\
M_a &=&  \frac{1}{2} c_a \frac{F_X}{\Lambda_{\rm IR}} .   
 \label{softmass}
\end{eqnarray}
For Higgs superfields, we can generally introduce 
 contact terms between $X$ and the gauge invariant product
 of up- and down-type Higgs superfields, $(H_u H_d)$. 
Such terms induce $\mu$-term and $B$-parameter 
 of the order of $F_X/\Lambda_{\rm IR}$ 
 through the Giudice-Masiero mechanism \cite{GM}. 
Now, since the scale of the gravity mediation is warped down 
 to the low scale, 
 the ``low scale gravity mediation'' has been realized.

For completeness, here we present a concrete model 
 of the hidden sector as an example. 
When we discuss the SUSY breaking mechanism 
 in extra dimension models, 
 the radion field is generally involved 
 and a mechanism to stabilize the extra dimensional radius 
 is strongly related to the SUSY breaking mediation. 
In the supersymmetric warped extra dimension scenario, 
 several ways to stabilize the radius 
 have been proposed \cite{SUSYL1, GLN, GNO, MO}. 
A model proposed in \cite{MO} is remarkable for our aim, 
 because the radius is stabilized in supersymmetric way 
 in the model, and the resultant supersymmetric radion mass 
 is so heavy that the radion potential is little affected by 
 the SUSY breaking on a brane. 
Here, we assume such a radius stabilization mechanism 
 by which the vacuum expectation value (VEV) of the radion 
 is completely fixed almost independently of 
 the SUSY breaking mechanism on a brane. 
Then, $\omega$ is dealt with as a constant in the following. 

We present a simple Lagrangian for the chiral superfield ($X$) 
 in the hidden sector on the IR brane such that 
\begin{eqnarray} 
 \int d^4 \theta \; 
 \omega^\dagger \omega  \; X^\dagger X 
 + \left( 
   \int d^2 \theta \; \omega^3 m^2 X + {\rm h.c.} 
 \right) ,
\end{eqnarray} 
where $m$ is a mass parameter. 
This is nothing but the Polonyi model \cite{Polonyi} on the IR brane. 
Rescaling $X$ to give the canonical Kahler potential, 
 $X \rightarrow  X/\omega $, 
 we obtain the SUSY breaking (non-zero $F$-term of $X$) 
 in four dimensional effective theory, 
\begin{eqnarray} 
 F_X = \left( \omega m \right)^2 . 
\end{eqnarray} 
The SUSY breaking scale is controlled by 
 the mass parameter $m$ accompanied 
 by the warp factor $\omega$ as expected.  
Depending on the value of $\Lambda_{\rm IR}$, 
 we take a suitable value for the parameter $m$ in the superpotential 
 so as to provide the typical soft mass scale around the electroweak scale. 
Only with the canonical Kahler potential, 
 there is a pseudo-flat direction in the scalar potential 
 and VEV of $X$ is undetermined. 
A simple way to lift up the pseudo-flat direction is 
 to introduce higher order terms in the Kahler potential. 
When we simply add a term, $- c (X^\dagger X)^2/M_5^2$,  
 with a dimensionless coefficient $c > 0$, 
 the potential minimum is realized at $\langle X \rangle=0$. 
In this simple case, 
 mass of the hidden sector scalar is given by 
 $m_X = 2 \sqrt{c} F_X/\Lambda_{\rm IR}$, 
 which is the same order of the soft SUSY breaking mass scale 
 in the visible sector.

Vacuum energy (cosmological constant) in supergravity 
 has two contributions: 
 One is positive from the SUSY breaking 
 and the other is negative from VEV of the superpotential 
 which couples to the compensating multiplet. 
This negative contribution is the result from 
 the fact that the Kahler potential 
 of the compensating multiplet has a wrong sign 
 in Eq.~(\ref{pureSUGRA}). 
To obtain the vanishing (almost zero) cosmological constant, 
 we simply put a constant superpotential on the UV brane, $ W_{\rm UV} $. 
 From Eqs.~(\ref{pureSUGRA}) and (\ref{UVLagrangian}), 
 total vacuum energy is described as 
\begin{eqnarray} 
E_{\rm vac} \simeq |F_X|^2 -3 \frac{|W_{\rm UV}|^2}{M_P^2} \simeq 0 ,  
 \label{vacuumenergy} 
\end{eqnarray} 
where 
$M_P^2 = M_5^3/k$ as mentioned above, and
 the constant superpotential $W_{\rm UV}$ has been tuned 
 so as to cancel out the positive contribution 
 from the SUSY breaking.

Gravity multiplet resides in the bulk, 
 whose zero-mode represents the gravity multiplet 
 in effective four dimensional supergravity. 
Since the gravity sector in effective four dimensional theory 
 should be reproduced correctly, 
 we obtain the usual formula for the gravitino mass 
 in four dimensional supergravity, 
 $m_{3/2} \simeq W_{\rm UV}/M_P^2$. 
Considering the condition of the vanishing cosmological constant, 
 the gravitino mass is usually expressed as 
\begin{eqnarray} 
m_{3/2} \simeq \frac{W_{\rm UV}}{M_P^2} 
 \simeq \frac{F_X}{M_P}.  
 \label{gravitinomass}
\end{eqnarray} 
In our scenario, 
 the scale of the gravity mediation is warped down 
 and the typical soft mass scale is given 
 by $\tilde{m} \simeq F_X/\Lambda_{\rm IR}$, 
 so that the gravitino mass is further rewritten as 
\begin{eqnarray}
 m_{3/2} \simeq \frac{F_X}{M_P} \simeq 
 \tilde{m} \times 
 \left( \frac{\Lambda_{\rm IR} }{M_P }    \right).  
 \label{gravitinomass2}
\end{eqnarray} 
Therefore, in the above setup, 
 the gravitino is naturally the LSP, 
 because of the suppression factor $\Lambda_{\rm IR}/M_P$. 
Similar result has been discussed 
 in the flat extra dimension model \cite{BHK}, 
 where $\Lambda_{\rm IR}$ is replaced by 
 $M_5$ smaller than $M_P$. 
We can reproduce this result by setting $M_5 < M_P$ 
 and taking the flat space-time limit $ k \rightarrow 0$.

\section{Phenomenology of low scale gravity mediation}

As discussed in the previous section, 
 the IR cutoff, $\Lambda_{\rm IR}$, is the model parameter, 
 and we can take any values for it. 
Accordingly, the SUSY breaking scale should be suitably 
 chosen so as to provide the typical soft mass scale 
 around the electroweak scale. 
In this section, we address phenomenologies 
 of the low scale gravity mediation scenario 
 for various IR cutoff scales.

\subsection{Phenomenology with the LSP gravitino}

As shown in the previous section, 
 the gravitino is naturally the LSP 
 due to the suppression factor $\Lambda_{\rm IR}/M_P$ 
 in Eq.~(\ref{gravitinomass2}), 
 so that it can be a candidate of the dark matter 
 in the present universe. 
Since there is no such a suppression factor 
 in the conventional minimal supergravity scenario, 
 the gravitino mass is normally of the same order of 
 the typical soft mass scale and 
 the gravitino is not so likely to be the LSP. 
Again, note that, in the warped extra dimension scenario, 
 we can take any values for $\Lambda_{\rm IR}$ 
 without serious fine-tuning among the original parameters 
 in the gravity sector, $M_5$, $k$ and $r$. 
Therefore, we can consider a wide range of 
 the LSP gravitino mass according to values of the warp factor. 
This feature is similar to the GMSB scenario, 
 where the gravitino mass varies with the messenger scale. 
The crucial difference is that 
 our model, as the same as the minimal supergravity scenario, 
 has more flexibility for sparticle mass spectrum 
 than the one in the GMSB scenario.

Since couplings among the gravitino, particles and sparticles 
 in the MSSM are suppressed by the Planck mass, 
 the gravitino cannot be in thermal equilibrium 
 in the early universe. 
There are two generic ways through which LSP gravitinos 
 are produced in the early universe. 
One is thermal production 
 through scattering and decay processes of the MSSM particles 
 in thermal plasma. 
In this case, the relic density of the gravitino 
 is evaluated as \cite{TP}
\begin{eqnarray} 
\Omega^{\rm TP} h^2 \sim 0.2 
 \left( \frac{T_R}{10^{10} {\rm GeV}} \right ) 
 \left( \frac{100 {\rm GeV}}{m_{3/2}} \right ) 
 \left( \frac{M_3}{1 {\rm TeV}} \right )^2 ,  
\end{eqnarray} 
where $T_R$ is the reheating temperature after inflation 
(which should be smaller than $\Lambda_{\rm IR}$ 
 due to theoretical consistency), 
 and $M_3$ is the running gluino mass. 
The other is non-thermal production through 
 the late time decay of a quasi-stable next LSP 
 after its decoupling from the thermal plasma 
 \cite{sWIMPcosmology}. 
In this case, the relic density of the LSP gravitino 
 is related to the relic density of the next LSP, 
\begin{eqnarray} 
 \Omega^{\rm NTP} h^2 
 = \frac{m_{3/2}}{M_{\rm NLSP}} 
  \Omega_{\rm NLSP} h^2 , 
\end{eqnarray} 
where $\Omega_{\rm NLSP} h^2$ would be the relic density 
 of the next LSP if it were stable, 
 and $M_{\rm NLSP}$ denotes its mass. 
By appropriately fixing the gravitino mass, 
 the reheating temperature and sparticle mass spectrum, 
 the relic density suitable for the dark matter 
 can be obtained. 
However, some cosmological constraints should be considered as well 
 \cite{BBNetc}. 
Since the gravitino couples to the MSSM particles 
 very weakly, the next LSP decays into the LSP gravitino 
 and standard model particles at late time. 
If it decays after big bang nucleosynthesis (BBN), 
 its energetic daughters would destroy light nuclei 
 through photo- and hadro-dissociation 
 and, as a result, upset the successful prediction of BBN.  
Furthermore, late time injection of energetic photons 
 produced by the next LSP decay would distort the spectrum 
 of the observed cosmic microwave background. 
These considerations will constrain the model parameters 
 to consistently realize the gravitino dark matter scenario. 

The quasi-stable next LSP opens up 
 an interesting possibility in collider physics. 
The decay rate of the next LSP ($\tilde{\Psi}$) 
 into a standard model particle ($\Psi$) 
 and the LSP gravitino ($\psi_{3/2}$) is given by 
\begin{eqnarray}
 \Gamma (\tilde{\Psi} \rightarrow \Psi \psi_{3/2}) 
 = \frac{\kappa m_{\tilde{\Psi}}^5}{48 \pi M_P^2 m_{3/2}^2} 
  \left( 1- \frac{m_\Psi^2}{m_{\tilde{\Psi}}^2}
  \right)^4 , 
\end{eqnarray}
where $\kappa \sim 1 $ is a model-dependent mixing parameter 
 among superpartners and standard model particles. 
Thus, the life time of the next LSP is estimated as 
\begin{eqnarray}
 \tau_{\tilde{\Psi}} \sim 
 10^8 {\rm sec}  \times 
 \left( \frac{100 {\rm GeV}}{m_{\tilde{\Psi}}} \right)^5 
 \left( \frac{m_{3/2}}{100 {\rm GeV}} \right)^2 
 \sim 
 10^8 {\rm sec}  \times 
 \left( \frac{100 {\rm GeV}}{\tilde{m}} \right)^3 
 \left( \frac{\Lambda_{\rm IR}}{M_P} \right)^2 . 
\end{eqnarray}
Here, in the last equality, we have replaced the mass 
 of the next LSP ($\tilde{m}_{\Psi}$) 
 into the typical sparticle mass and used Eq.~(\ref{gravitinomass2}). 
If $\Lambda_{\rm IR} \gg 10^{10}$ GeV, 
 the decay length well exceeds the detector size 
 of the LHC and the ILC, 
 and the next LSP decay takes place outside the detector. 
In this case, there have been interesting proposals \cite{sWIMPcollider}
 for the way to trap quasi-stable next LSPs outside the detector, 
 when the next LSP is a charged particle. 
Detailed studies of the next LSP decay may provide 
 precise measurements of the gravitino mass 
 and the four dimensional Planck mass. 
On the other hand, if $\Lambda_{\rm IR} \ll 10^{10}$ GeV, 
 the next LSP decays within the detector. 
In the GMSB scenario 
 where the next LSP can be neutralino and 
 right-handed slepton \cite{DTW}, 
 it has been pointed out \cite{colliderGMSB, DTW} 
 that the next LSP decay provides very characteristic 
 SUSY signatures 
 with leptons and/or photons accompanied by missing $E_T$. 
For detailed studies on general types of the next LSP, 
 see Ref. \cite{Tevatron}. 
Our model can naturally provide such general cases, 
 since it has more flexibility for sparticle mass spectra 
 than those in the GMSB scenario.

\subsection{Collider phenomenology involving the hidden sector field} 

%
As mentioned above, 
 we can take $\Lambda_{\rm IR} ={\cal O}$(1 TeV) 
 without any serious fine-tuning for the original model parameters. 
If the effective cutoff scale is low enough, 
 higher dimensional interactions suppressed 
 by $\Lambda_{\rm IR}$ have an impact on collider physics. 
Note that the contact operators relevant to the gravity mediation  
 also provide contact interactions 
 between the hidden sector scalar field and the standard model fields. 
 From the operator giving masses to gauginos in Eq.~(\ref{contactOP}), 
 we can extract interactions among 
 the hidden sector scalar and the standard model gauge bosons 
 such that 
\begin{eqnarray}
 {\cal L}_{\rm int} = 
 - \frac{1}{4} \int d^2 \theta \; 
  c_a \frac{X}{\Lambda_{\rm IR}} {\cal W}^{a \alpha}{\cal W}^a_\alpha 
 \supset 
  -\frac{c_a}{4 \sqrt{2}} \frac{\chi}{\Lambda_{\rm IR}} 
  {\cal F}^{a \mu \nu}{\cal F}^a_{\mu \nu}  
  - \frac{c_a}{8 \sqrt{2}} \frac{a}{\Lambda_{\rm IR}} 
  {\cal F}^{a \mu \nu}\tilde{{\cal F}}^a_{\mu \nu} ,  
 \label{intgauge}
\end{eqnarray}
where we have decomposed the hidden scalar field $X$ 
 into two real scalar fields, $X= (\chi +i a )/\sqrt{2}$, 
 and ${\cal F}^a$ and $\tilde{\cal F}^a$ are the field strength 
 and its dual of corresponding standard model gauge fields, 
 respectively. 
In the case of $\langle X \rangle \ll \Lambda_{\rm IR}$, 
 the operators in Eq.~(\ref{contactOP}) 
 also give interactions between 
 the hidden scalar and standard model fermions, 
\begin{eqnarray}
 {\cal L}_{\rm int} = 
  \int d^4 \theta \; 
  c_A \frac{X + X^\dagger}{\Lambda_{\rm IR}} Q_i^\dagger Q_i 
 \supset 
  \sqrt{2} c_A \frac{\chi}{\Lambda_{\rm IR}} 
  {\cal L}^{\rm fermion}_{\rm kin},    
 \label{intfermion}
\end{eqnarray}  
where ${\cal L}^{\rm fermion}_{\rm kin}$ is the kinetic term 
 for each standard model fermion.

Now we investigate collider phenomenologies involving 
 the hidden scalar based on the above interactions. 
In the following, to make our discussion clear, 
 we do not specify a concrete potential 
 of the hidden sector fields, $\chi$ and $a$, 
 so that their masses are dealt with as free parameters. 
Furthermore, we concentrate on the phenomenology 
 involving only $\chi$, the real part of $X$, for simplicity. 
The general case involving both $\chi$ and $a$ 
 can be investigated in the same way, and 
 we will arrive at almost the same conclusions.

Let us begin with phenomenology at the LHC. 
If $\chi$ is light enough and $\Lambda_{\rm IR}$ are low enough, 
 it may be possible to produce the hidden scalar 
 at the collider through the interactions 
 in Eq.~({\ref{intgauge}}). 
For $c_1 \simeq c_2 \simeq c_3 \simeq 1$, 
 the dominant $\chi$ production process at the LHC 
 is the gluon fusion process. 
Note that the dominant production process 
 of the Higgs boson of the standard model 
 is the same gluon fusion process 
 through the effective interaction 
 among the Higgs boson ($h$) and gluons 
 induced by top quark one-loop diagram \cite{HHG}, 
\begin{eqnarray}
 {\cal L}_{\rm eff} = 
  - \frac{\alpha_s}{16 \pi} F_{1/2}(\tau_t) 
    \frac{h}{v} G^{a \mu \nu} G^a_{\mu \nu}.  
 \label{hgg} 
\end{eqnarray}
Here, $F_{1/2}$ is the form factor given as 
\begin{eqnarray}
F_{1/2} = -2 \tau 
 \left[ 1+ \left( 1 - \tau  \right) f(\tau) \right] ,
\end{eqnarray}
where 
\begin{eqnarray}
f(\tau)  =  \left\{ \begin{array}{cc}
      \left[ \sin^{-1}\left( 1/\sqrt{\tau}\right)\right]^2 &
               ({\rm for}~\tau\ge 1), \\
      -\frac{1}{4} 
    \left[ \ln \left(
       \frac{1+\sqrt{1-\tau}}{1-\sqrt{1-\tau}} \right)-i \pi \right]^2 &
               ({\rm for}~\tau < 1), 
                      \end{array}  \right.  
\end{eqnarray}
 $\tau_t=4 m_t^2/q^2$ with momentum transfer $q^2$ 
 in the direction to the Higgs boson, 
 and $v=246$ GeV is VEV of Higgs field. 
Interestingly, the effective interaction is of the same form 
 as the one in Eq.~(\ref{intgauge}). 
Therefore, in the production process, the hidden scalar $\chi$ 
 behaves like the Higgs boson in the standard model.  
Comparing coefficients of their interactions, 
 we find that their production cross sections become comparable 
 for $\Lambda_{\rm IR} \simeq 10$ TeV, 
 assuming the same masses for them.

When we consider the decay process of $\chi$, 
 we find a big difference between $\chi$ and the Higgs boson. 
 From Eqs.~(\ref{intgauge}) and (\ref{intfermion}), 
 the partial decay width of $\chi$ into the standard model 
 gauge bosons and fermions is easily calculated. 
The decay width into a pair of gauge bosons 
 are found to be 
\begin{eqnarray} 
\Gamma (\chi \rightarrow g g )
&=& \frac{c_3^2}{16 \pi} 
    \frac{m_\chi^3}{\Lambda_{\rm IR}^2},  
 \nonumber \\ 
\Gamma (\chi \rightarrow \gamma \gamma)
&=& \frac{ \left(
  c_1 \cos^2 \theta_w +c_2 \sin^2 \theta_w \right)^2}{128 \pi} 
    \frac{m_\chi^3}{\Lambda_{\rm IR}^2} , 
 \nonumber \\ 
\Gamma (\chi \rightarrow Z Z)
&=& \frac{
 \left( c_1 \sin^2 \theta_w +c_2 \cos^2 \theta_w \right)^2}{1024 \pi} 
 \frac{m_\chi^3}{\Lambda_{\rm IR}^2} 
 \; \beta_Z \left(
 3 + 2 \beta_Z^2 +3 \beta_Z^4\right) ,
 \nonumber \\
\Gamma (\chi \rightarrow W W)
 &=& \frac{c_2^2}{512 \pi} 
 \frac{m_\chi^3}{\Lambda_{\rm IR}^2} 
 \; \beta_W \left( 
 3 + 2 \beta_W^2 +3 \beta_W^4 \right) ,
 \nonumber \\ 
\Gamma (\chi  \rightarrow \gamma Z)
 &=& \frac{ 
 (c_1-c_2)^2  \sin^2 \theta_w \cos^2 \theta_w}{64 \pi} 
  \frac{m_\chi^3}{\Lambda_{\rm IR}^2} 
 \left(  1- \frac{m_Z^2}{m_\chi^2}  \right)^3, 
\label{width-gauge} 
\end{eqnarray} 
where $m_\chi$ is the mass of the hidden scalar $\chi$, 
 $\theta_w$ is the weak mixing angle, 
 $\beta_Z = \sqrt{1-4 (m_Z/m_\chi)^2} $, 
 and $\beta_W = \sqrt{1-4 (m_W/m_\chi)^2}$.  
The interaction of Eq.~(\ref{intfermion}) gives 
 the partial decay width into a fermion pair,  
\begin{eqnarray} 
 \Gamma (\chi  \rightarrow f \bar{f} )
 = \frac{c_A^2}{4 \pi} 
  \frac{m_f^2  m_\chi}{\Lambda_{\rm IR}^2} \; 
 \beta_f^3 \times N_c, 
\end{eqnarray} 
where $m_f$ is the mass of the final state fermions, 
     $\beta_f= \sqrt{1- 4 ( m_f/m_\chi)^2} $, 
 and $N_c$ is the color factor for the final state fermions. 
Since fermions couple with  $\chi$ through their kinetic terms, 
 the decay width is proportional to $m_f^2$, 
 so that decay channels into light fermions are very much 
 suppressed compared to those into gauge boson pairs. 
This result contrasts with the fact that 
 the dominant decay channel of the light Higgs boson 
 with mass $m_h < 2 m_W$ is into bottom and anti-bottom quarks, 
 since the Higgs boson decay into gauge bosons 
 occurs through one-loop radiative corrections. 

\begin{figure}
\begin{center}
\leavevmode
  \scalebox{1.2}{\includegraphics*{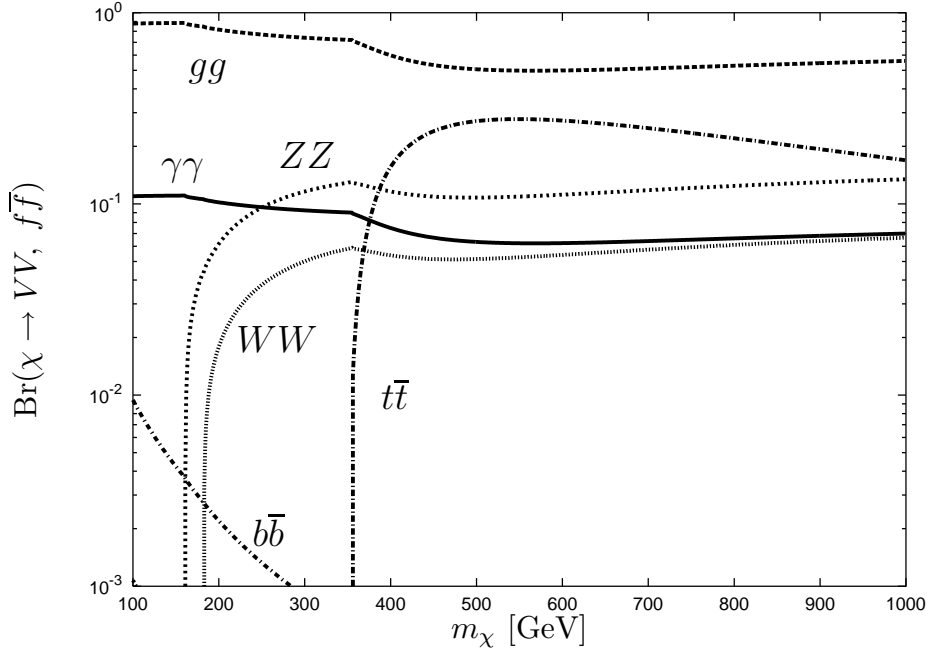}}
\caption{
 The branching ratio of the hidden scalar ($\chi$) 
 as a function of its mass $m_\chi$ for $c_1=c_2=c_3=c_A=1$. 
 The plot on the lower-left corner corresponds 
 to $Br(\chi \rightarrow \tau \bar{\tau})$. 
}
\end{center}
\end{figure}

We show the branching ratio of the $\chi$ decay in Fig. 1. 
Here, we have considered only the decay channels into 
 the standard model particles, 
 assuming that all the sparticles and Higgs bosons in the MSSM 
 are heavier than $\chi$. 
If $\chi$ is heavy enough, 
 it can decay into sparticle pairs and Higgs boson pairs. 
Their interactions are found to be similar to Eq.~(\ref{intfermion}), 
 and the partial decay width into sparticle and Higgs boson pairs 
 is proportional to the mass of the final states. 
We see that the branching ratio of the $\chi$ decay 
 is quite different from that of the Higgs boson. 
In particular, the branching ratio of 
 $\chi \rightarrow \gamma \gamma$ is large, 
 $Br(\chi \rightarrow \gamma \gamma) \simeq 0.1$. 
On the other hand, 
 the branching ratio of the Higgs boson 
 into two photons is at most $10^{-3}$, 
 even when the Higgs mass is light $m_h <  2 m_W $. 
This fact implies that 
 once $\chi$ is produced at the LHC, 
 the signature of $\chi$ is distinguishable 
 from the Higgs boson one.

In the MSSM, the lightest Higgs boson is like the standard model 
 Higgs boson and its mass is too light to decay 
 into weak gauge boson pairs. 
The most important channel 
 for the lightest Higgs boson search at the LHC 
 is its decay process into two photons. 
Therefore, the $\chi$ production and its decay process 
 into two photons have a great impact on 
 the (lightest SUSY) Higgs boson search at the LHC. 
To see this, let us evaluate a ratio between two photon events 
 from the $\chi$ decay and the Higgs boson decay. 
The ratio of $\chi$ and the Higgs boson production rates 
 can be estimated from the ratio between the coefficients of 
 the ${\cal F}^{a \mu \nu}{\cal F}^a_{\mu \nu}$ terms 
 in Eqs.~(\ref{intgauge}) and (\ref{hgg}). 
For $m_\chi = m_h= 120$ GeV and $c_1=c_2=c_3=c_A=1$, 
 the event number ratio as a function of the effective cutoff scale 
 is depicted in Fig. 2.
\begin{figure}
\begin{center}
\leavevmode
  \scalebox{1.2}{\includegraphics*{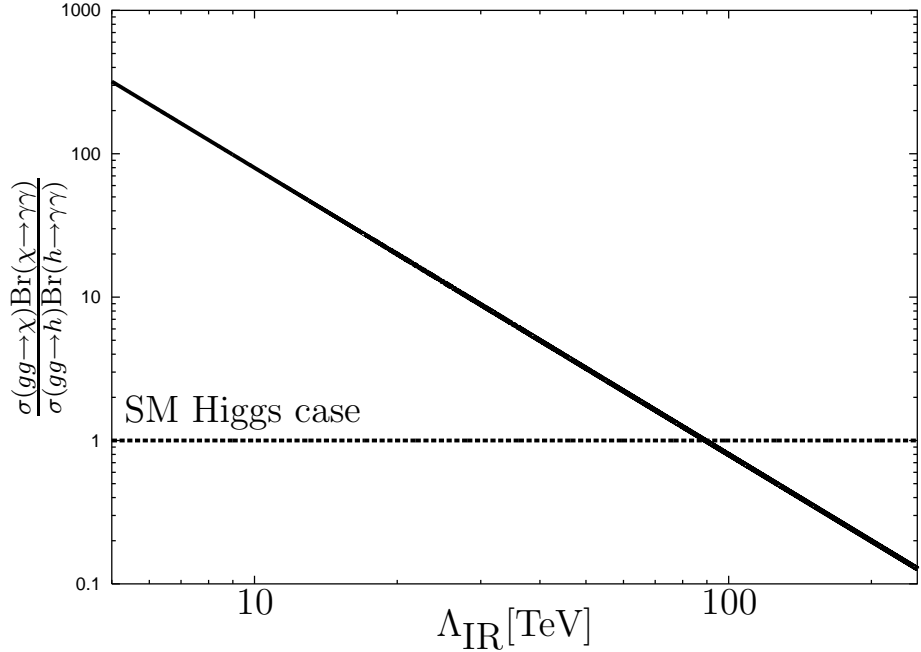}}
\caption{
The ratio between two photon events 
 from the $\chi$ production and the Higgs boson production 
 at the LHC, as a function of $\Lambda_{\rm IR}$, 
 for $c_1=c_2=c_3=c_A=1$ and $m_\chi=m_h=120$ GeV. 
}
\end{center}
\end{figure} 
We can see that, for $\Lambda_{\rm IR} = 10$ TeV, 
 the number of events from $\chi$ production 
 is two order of magnitude larger than 
 that from the Higgs boson production! 
Even for $\Lambda_{\rm IR} = {\cal O}$(100 TeV), 
 the ratio is still of order one.  
Therefore, if $\Lambda_{\rm IR}$ is around 10 TeV, 
 the hidden sector scalar $\chi$ 
 can be discovered at the LHC with a very clean signature.

We discuss more on interesting features 
 of the low scale gravity mediation scenario. 
Note that there is one-to-one correspondence 
 between gaugino masses and the partial decay width 
 of the hidden sector scalar into gauge boson pairs, 
 because they are originated from the same contact operators. 
Considering that 
 the quantity $M_a/\alpha_a$ is invariant 
 under renormalization group equations \cite{RGE}, 
 the ratio between gaugino masses at the typical soft mass scale 
 $\tilde{m}$ is given by 
\begin{eqnarray}
M_1: M_2 : M_3= 
 c_1 \frac{\alpha_1(\tilde{m})}{\alpha_1(\Lambda_{\rm IR})}:
 c_2 \frac{\alpha_2(\tilde{m})}{\alpha_2(\Lambda_{\rm IR})}:
 c_3 \frac{\alpha_3(\tilde{m})}{\alpha_3(\Lambda_{\rm IR})},   
\end{eqnarray}
which is determined by the ratio between $c_a$. 
As shown in Eq.~(\ref{width-gauge}), 
 the ratio between the partial decay width into 
 pairs of gauge bosons is also 
 fixed by the ratio between $c_a$. 
Therefore, once gauginos and the hidden sector scalar 
 are discovered at future colliders 
 and their masses and the partial decay width of the hidden scalar 
 are precisely measured, 
 we can check the origin of SUSY breaking mediation 
 by examining this one-to-one correspondence.

Finally, let us investigate phenomenology at the ILC. 
The ILC, the liner $e^+ e^-$ collider, is  
 the so-called Higgs boson factory
 where a large number of Higgs bosons will be produced. 
The most clean channel of the Higgs boson production at the ILC 
 is the associated Higgs production (Higgsstrahlung production), 
 $e^+ e^- \rightarrow Z h$, 
 through the standard model interaction 
 ${\cal L}_{\rm int}= \frac{m_Z^2}{v} h Z^\mu Z_\mu$. 
Since the hidden sector scalar $\chi$ has the vertex 
 among $Z$-boson in Eq.~(\ref{intgauge}), 
 we can consider the same associated process 
 for the $\chi$ production at the ILC%
\footnote{
 For $\chi$ productions, 
 we can also consider the process $e^+ e^- \rightarrow \gamma \chi$ 
 as one of main production processes, 
 while such a process is negligible for the Higgs boson production.
 Studies on this process itself would be interesting. }. 
In the case of the universal couplings, $c_1=c_2=c_3$, 
 for simplicity, 
 the cross section of the process 
 $e^+ e^- \rightarrow Z \chi$ is found to be%
\footnote{
In the general case $c_1 \neq c_2$, 
 the process 
 $ e^+ e^- \rightarrow \gamma^* \rightarrow Z \chi$ 
 should be included. } 
\begin{eqnarray} 
& & \frac{d \sigma}{d \cos \theta} (e^+e^- \rightarrow Z \chi)
 =  \frac{1}{64 \pi s}  \sqrt{\frac{E_Z^2-m_Z^2}{s}} 
\nonumber \\
& \times & 
 \frac{c_2^2}{2}
 \left(  \frac{e}{\sin \theta_w \cos \theta_w} \right)^2 
 \left( g_L^2+ g_R^2 \right) 
 \left( \frac{s}{s-m_Z^2}   \right)^2 
 \frac{E_Z^2}{\Lambda_{\rm IR}^2}  
 \left( 
 1 + \cos^2 \theta + \frac{m_Z^2}{E_Z^2} \sin^2 \theta 
 \right) , 
\end{eqnarray} 
where $\cos \theta$ is the scattering angle of 
 the final state $Z$-boson, 
 $g_L=-1/2+ \sin^2 \theta_w$, $g_R= \sin^2 \theta_w$, and 
 $E_Z= \frac{\sqrt{s}}{2} \left( 1+ \frac{m_Z^2 - m_\chi^2}{s} \right) $. 
\begin{figure}
\begin{center}
\leavevmode
  \scalebox{1.2}{\includegraphics*{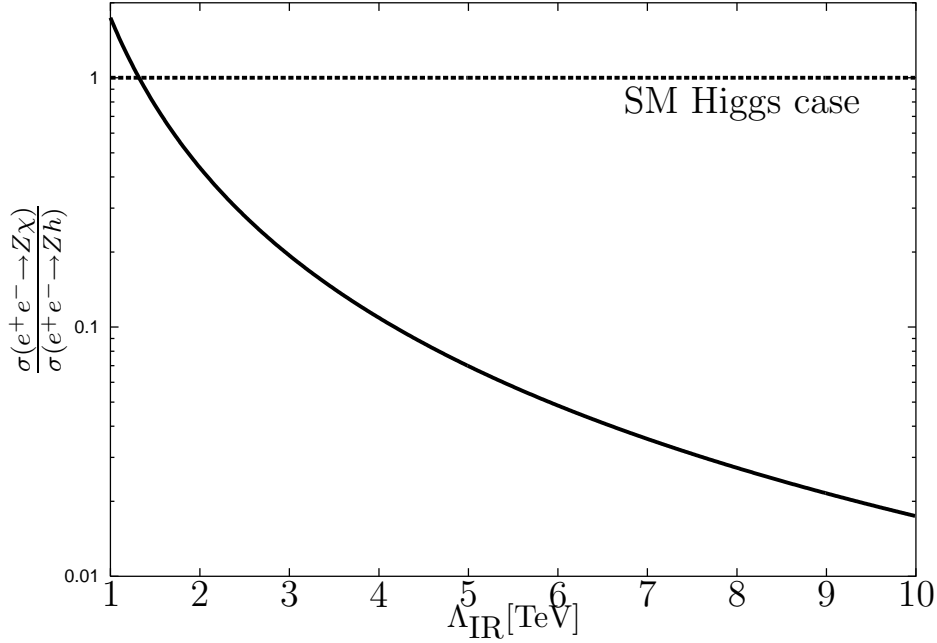}}
\caption{
The ratio of total cross sections 
 between the associated $\chi$ and Higgs productions 
 as a function of $\Lambda_{\rm IR}$, 
 at the ILC with the collider energy $\sqrt{s} = 1$ TeV. 
Here, we have fixed the parameters 
 such as 
 $m_\chi=m_h=120$ GeV and $c_1=c_2=c_3= c_A=1$. 
The ratio becomes one for $\Lambda_{\rm IR} \simeq 1.3$ TeV. 
}
\label{fig:Fig3}
\end{center}
\end{figure}
Since Higgs boson couples to a pair of $Z$-boson at tree level, 
 its production cross section is mostly larger 
 than the one of $\chi$ production. 
In Fig.~3, we show the ratio of the total cross sections 
 between $\chi$ and Higgs boson productions 
 as a function of $\Lambda_{\rm IR}$ 
 at the ILC with the collider energy $\sqrt{s} = 1$ TeV. 
The ratio, 
 $\sigma(e^+e^-\rightarrow Z \chi)/\sigma(e^+e^-\rightarrow Z h)$, 
 becomes one for $\Lambda_{\rm IR} \simeq 1.3$ TeV, 
 and it decreases proportionally to $1/\Lambda_{\rm IR}^2$.

The coupling manner among $\chi$ and the $Z$-boson pair 
 is different from that of the Higgs boson, 
 and this fact reflects into the difference 
 of the angular distribution of the final state $Z$-boson. 
\begin{figure}
\begin{center}
\leavevmode
  \scalebox{1.2}{\includegraphics*{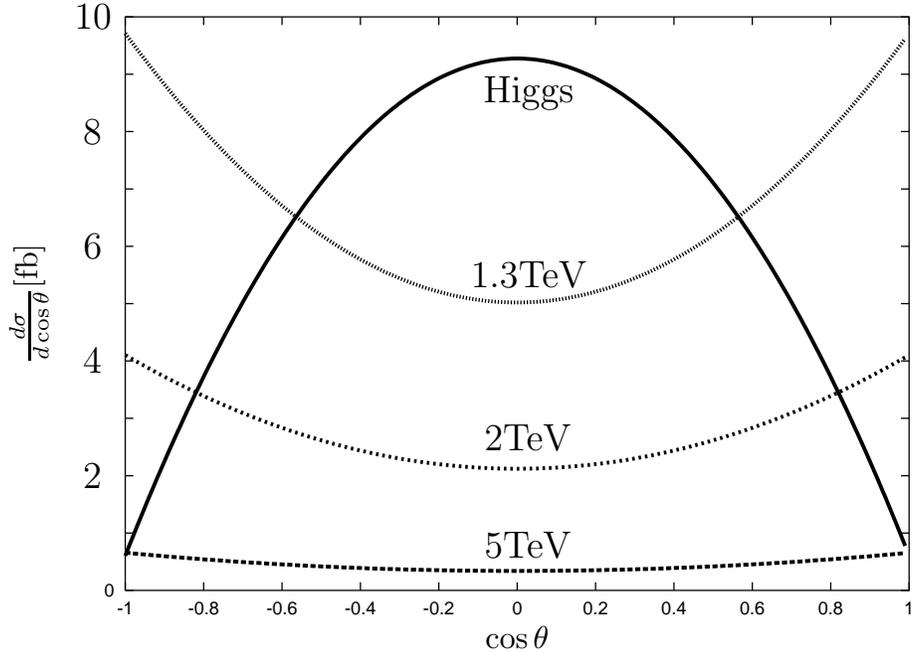}}
\caption{
The angular dependence of the cross sections 
 for $m_\chi=m_h=120$ GeV and $c_1=c_2=c_3= c_A=1$, 
 at the ILC with the collider energy $\sqrt{s} = 1$ TeV. 
The standard model Higgs boson case is depicted as the solid line, 
 while the others correspond to the $\chi$ productions 
 with $\Lambda_{\rm IR} =1.3$, $2$ and $5$ TeV, respectively, 
 from above. 
}
\label{fig:Fig4}
\vspace{-0.5cm}
\end{center}
\end{figure}
In the high energy limit, 
 we find 
 $\frac{d \sigma}{d \cos \theta} (e^+e^- \rightarrow Z \chi) 
   \propto 1+ \cos^2 \theta$, 
 while $\frac{d \sigma}{d \cos \theta} (e^+e^- \rightarrow Z h) 
   \propto 1- \cos^2 \theta$. 
Fig.~4 shows the angular distributions 
 of the associated $\chi$ and Higgs boson productions, 
 respectively. 
Even if $m_\chi = m_h$ and the cross sections 
 of $\chi$ and Higgs boson productions are comparable, 
 the angular dependence of the cross section 
 can distinguish the $\chi$ production from the Higgs boson one. 
Of course, detecting two photons from the $\chi$ decay 
 with the sizable branching ratio 
 $Br(\chi \rightarrow \gamma \gamma) \sim 0.1$ 
 would be an easy way to distinguish $\chi$ from Higgs boson, 
 as discussed in the case of the LHC.

\section{Conclusions and discussions} 
We have proposed the low scale gravity mediation scenario 
 with the warped extra dimension. 
The setup of the scenario is that 
 both of the hidden and visible sectors co-exist on the IR brane. 
This setup is the same as 
 in the four dimensional minimal supergravity scenario 
 except for the gravity multiplet residing in the bulk. 
We have considered the gravity mediated SUSY breaking 
 through the contact operators 
 between the hidden and the visible sector superfields. 
The crucial point is that 
 the effective cutoff on the IR brane is warped down, 
 so that the gravity mediation takes place at low energies. 
As a result, the gravitino is naturally the LSP, 
 just as in the GMSB scenario. 
However, 
 our gravity mediated scenario has more flexibility 
 for sparticle mass spectra 
 than those in the GMSB scenario. 
We have briefly discussed phenomenologies 
 related to the LSP gravitino scenario.

If the effective cutoff scale is low enough, 
 for example, $\Lambda_{\rm IR} ={\cal O}$(10 TeV), 
 our scenario provides interesting phenomenologies 
 at the future colliders. 
The contact operators relevant to the gravity mediation 
 also provide contact interactions 
 among the hidden sector scalars 
 and the standard model particles. 
We have investigated collider physics 
 involving the hidden sector scalar fields 
 at the LHC and the ILC. 
Interestingly, the hidden sector scalars 
 behave like the standard model Higgs boson 
 in their production processes and, therefore, 
 the existence of such scalars has an great impact 
 on the Higgs boson search at the colliders. 
Since the decay process of the hidden scalars 
 is quite different from the Higgs boson one, 
 and, once produced, they will provide us 
 with a very clean signature. 
The hidden sector may be no longer hidden.

Several discussions are in order. 

In this paper, we have concentrated our discussion 
 only on the contact operators relevant to the gravity mediation.
In general, we may introduce contact operators 
 among the visible sector fields themselves, 
 which induce contact interactions 
 among the standard model particles. 
For such contact interactions,  
 the lower bound on $\Lambda_{\rm IR}$ by the current experiments 
 should be taken into account. 
The electroweak precision measurements 
 give the lower bound, 
 $\Lambda_{\rm IR} \geq 5$ TeV \cite{EWPM}. 
If contact operators which cause FCNC processes 
 are considered, rough estimation gives a severer bound, 
 $\Lambda_{\rm IR} \geq 100$ TeV. 
We may expect that the severely constrained operators 
 are forbidden by some underlying flavor symmetry 
 which justifies the minimal ansatz.

Next, if $\Lambda_{\rm IR} ={\cal O}$(10 TeV), 
 in other words, $ \omega \sim 10^{-14}$, 
 taken as in the previous section, 
 the gravitino mass becomes too small, $m_{3/2} \sim 10^{-3}-10^{-4}$ eV, 
 to account for the dark matter density 
 in the present universe. 
In this case, our model must be extended 
 so as to implement a suitable candidate for the cold dark matter. 
Among various possibilities, 
 we notice that, 
 in extra dimensional models, 
 there is more flexibility for the scale of the gravitino mass. 
In fact, as discussed in a series of papers 
 \cite{ANS1, ANS2, GLN, GNO}, 
 it is generally possible for the gravitino mass 
 to be even the Planck scale. 
An important feature is that the gravity multiplet 
 residing in the bulk couples to fields on both branes. 
Thus, when we introduce an additional hidden sector 
 on the UV brane, 
 the gravitino directly picks up the SUSY breaking on the UV brane 
 and becomes massive. 
If the SUSY breaking scale is much larger 
 than the effective SUSY breaking scale on the IR brane, 
 total vacuum energy in Eq.~(\ref{vacuumenergy}) is replaced into 
\begin{eqnarray} 
 E_{\rm vac} \simeq |F_Y|^2 -3 \frac{|W_{\rm UV}|^2}{M_P^2} \simeq 0,
\end{eqnarray} 
 where $F_Y$ is the large SUSY breaking on the UV brane, 
 so that the gravitino mass is dominantly induced 
 from this SUSY breaking, $m_{3/2} \sim F_Y/M_P \gg F_X/M_P$. 
On the other hand, the visible sector residing on 
 the IR brane cannot directly feels the SUSY breaking 
 on the UV brane, because two branes are spatially separated. 

We must consider some possibilities on the SUSY breaking mediation 
 from the UV brane to the IR brane over the bulk space. 
One is due to quantum corrections 
 through the supergravity multiplet in the bulk, 
 whose contribution is evaluated as \cite{GRSST} 
\begin{eqnarray}
 \Delta \tilde{m} \sim m_{3/2} \; \omega^2 . 
\end{eqnarray}
Even if $m_{3/2} \sim M_P$, 
 this is negligible compared to 
 the low scale gravity mediation on the IR brane 
 with the strong warp factor, $ \omega \sim 10^{-14}$. 
In supergravity, the SUSY breaking mediation 
 through the superconformal anomaly \cite{AMSB} 
 (anomaly mediation) always exists. 
In warped extra dimensional models, 
 the anomaly mediation contribution 
 on the IR brane can be characterized by \cite{ANS1} 
\begin{eqnarray}
 \Delta \tilde{m}_{\rm AMSB} \sim \frac{F_\omega}{\omega} .
\end{eqnarray}
This contribution highly depends on a mechanism 
 to stabilize the fifth dimension. 
For example, in models of the radius stabilization 
 proposed in \cite{GLN, GNO}, 
 we obtain 
 $ \Delta \tilde{m}_{\rm AMSB} \sim m_{3/2} \omega^n $ 
 with a model parameter $n$ of order one. 
The setup of these models is that  
 the visible sector resides on the IR brane 
 while the hidden sector resides only on the UV brane%
\footnote{
To be precise, 
 SUSY on the UV brane is explicitly broken, 
 nevertheless we obtain softly broken SUSY theory on the IR brane. 
 This is a scenario proposed in \cite{GLN, GNO}, 
 ``emergent supersymmetry''. }. 
Thus, it is easy to combine our model with these models 
 by introducing the hidden sector also on the IR brane. 
When we fix the model parameter appropriately, $n > 1$, for example, 
 we can realize the situation 
 that the gravity mediation on the IR brane gives 
 the dominant contribution to the soft SUSY breaking parameters. 
In this case, 
 the gravitino is much heavier than sparticles in the MSSM,  
 and the lightest neutralino can be the LSP 
 and the candidate of the cold dark matter as usual. 

Taking the flat space-time limit, $k \rightarrow 0$, 
 in our model, 
 we obtain the effective cutoff scale 
 as $\Lambda_{\rm IR} = M_5$. 
If we take $M_5$ to be much smaller than 
 the four dimensional Planck scale, 
 we can realize the low scale gravity mediation 
 without the warp factor. 
However, in this case, 
 the low scale cutoff, $M_5 \ll M_P$, implies $1/r \ll M_5$ 
 in order to correctly reproduce the four dimensional Planck scale 
 through the relation, $M_P^2 \sim M_5^3 r$. 
Thus, one may claim a hierarchy problem between $1/r \ll M_5$. 
In warped extra dimension scenario, 
 there is no such a hierarchy problem, 
 thanks to the warp factor. 
In the following, we will show that there exists 
 a theoretical lower bound on $M_5$ 
 in the flat extra dimension scenario, 
 even if we admit the hierarchy between $1/r \ll M_5$.

Using the condition of the vanishing cosmological constant 
 in Eq.~(\ref{vacuumenergy}) and the typical soft mass scale 
 $\tilde{m} \sim F_X/M_5$, we obtain the relation%
\footnote{
In the flat extra dimensional scenario, 
 there is no difference between superpotentials 
 on the IR and UV brane, 
 since the AdS curvature is zero and, thus, $\omega = \phi$.}, 
\begin{eqnarray} 
 W_{UV} \sim F_X M_P  \sim \tilde{m} M_5 M_P .   
\end{eqnarray} 
Since $M_5$ is the cutoff scale of the original theory, 
 the theoretical consistency, $W_{UV} \leq M_5^3$, 
 implies the lower bound on the scale $M_5$ such as 
\begin{eqnarray} 
  M_5 \geq \sqrt{\tilde{m} M_P} \sim 10^{10} \;  {\rm GeV} 
\end{eqnarray} 
for $\tilde{m} \simeq 100$ GeV. 
Therefore, we cannot take $M_5$ as low as 1 TeV. 
In the warped extra dimension models, 
 the above condition is replaced by 
 $M_5 \geq \sqrt{\omega \tilde{m} M_P}$. 
For any $M_5$ satisfying this condition, 
 we can realize the effective low scale cutoff 
 by the warp factor, $\Lambda_{\rm IR} = \omega M_5$. 
There is no lower bound on the effective cutoff
 in the theoretical point of view.

Finally, let us consider an issue related to 
 the scale of the SUSY breaking. 
We can express the SUSY breaking scale in terms of 
 the typical soft mass scale and the effective cutoff, 
 $ \sqrt{F_X} \sim \sqrt{\tilde{m} \Lambda_{\rm IR}} $. 
When the scale of the SUSY breaking mediation,
  $\Lambda_{\rm IR}$, is very high, 
 for example, $\Lambda_{\rm IR} = M_P$ 
 in the usual minimal supergravity scenario, 
 the hierarchy between $\sqrt{F_X} \ll \Lambda_{\rm IR}$ 
 is necessary to provide soft SUSY breaking masses 
 around the electroweak scale. 
How to generate such a hierarchy could be an important issue 
 when one constructs a concrete SUSY breaking model. 
Dynamical SUSY breaking \cite{DSB} is a remarkable possibility, 
 in which the SUSY breaking scale is controlled 
 by the dynamical scale of some strong interaction 
 induced through the dimensional transmutation. 
Thus, there is no problem on the hierarchy.

In the warped extra dimension scenario, 
 any original dimensional parameters on the IR brane 
 are warped down according to their mass dimensions 
 such as 
\begin{eqnarray}
& M_5 &\rightarrow \omega M_5 =\Lambda_{\rm IR},  \nonumber \\
& \sqrt{F_X} &\rightarrow \omega \sqrt{F_X}. 
\end{eqnarray}
As discussed before, we can realize, for example, 
 $\Lambda_{\rm IR} \sim 10$ TeV 
 only with the mild hierarchy, 
 $M_5 \sim k \sim M_P$ and $1/r \sim 0.1 M_P$. 
In the same way, if we introduce a mild hierarchy 
 for the original SUSY breaking scale, 
 $\sqrt{F_X} \sim 0.1 M_P$, 
 we obtain the effective SUSY breaking scale 
 such as $\sqrt{F_X} \sim 0.1 M_P \rightarrow \sqrt{F_X} \sim 1$ TeV. 
Then, the typical soft SUSY breaking mass scale 
 appears around the electroweak scale, $\tilde{m} \sim 100$ GeV. 
This result implies that, 
 in order to provide the correct electroweak scale, 
 we do not need to introduce any additional mass scales 
 except for the four dimensional Planck scale. 
For $\mu$-parameter in the Higgs sector of the MSSM, 
 we can follow the same manner. 
When a mildly hierarchical $\mu$-parameter 
 such as $\mu \sim 0.1 M_P$ is introduced 
 in the original superpotential on the IR brane, 
 it becomes a suitable scale, $\mu \sim 1$ TeV, 
 in effective four dimensional theory. 
Taking $\Lambda_{\rm IR} \leq 10$ TeV can make everything 
 go well only with the mild hierarchy, 
 and so it would be the most natural setting.

\noindent {\large \bf Acknowledgments}

We would like to thank Markus A. Luty, Sukanta Dutta, 
 Siew-Phang Ng, and Yasuhiro Okada for useful discussions. 
Also, we are grateful to Junichi Kanzaki 
 for his interests in this work and for discussions 
 on signatures of hidden sector fields 
 at the LHC. 
N.O. would like to thank the Elementary Particle Physics Group 
 at the University of Maryland, and especially Markus A. Luty, 
 for their hospitality during the completion of this work. 
He also would like to thank 
 the Theoretical Particle Physics Group of 
 the Bartol Research Institute, and especially Siew-Phang Ng, 
 for the hospitality during his stay. 
The work of N.O. is partly supported by 
 the Grant-in-Aid for Scientific Research in Japan 
 (\#15740164, \#18740170 and \#16081211).
The work of T.Y. is supported by SISSA.


\end{document}